\PassOptionsToPackage{unicode}{hyperref}
\PassOptionsToPackage{hyphens}{url}
\documentclass[
]{article}
\usepackage{amsmath,amssymb}

\usepackage[T1]{fontenc}
\usepackage[utf8]{inputenc}
\usepackage{lmodern}
\usepackage{iftex}
\ifPDFTeX
  \usepackage[T1]{fontenc}
  \usepackage[utf8]{inputenc}
  \usepackage{textcomp} % provide euro and other symbols
\else % if luatex or xetex
  \usepackage{unicode-math} % this also loads fontspec
  \defaultfontfeatures{Scale=MatchLowercase}
  \defaultfontfeatures[\rmfamily]{Ligatures=TeX,Scale=1}
\fi
\usepackage{lmodern}
\ifPDFTeX\else
\fi
\IfFileExists{upquote.sty}{\usepackage{upquote}}{}
\IfFileExists{microtype.sty}{% use microtype if available
  \usepackage[]{microtype}
  \UseMicrotypeSet[protrusion]{basicmath} % disable protrusion for tt fonts
}{}
\makeatletter
\@ifundefined{KOMAClassName}{% if non-KOMA class
  \IfFileExists{parskip.sty}{%
    \usepackage{parskip}
  }{% else
    \setlength{\parindent}{0pt}
    \setlength{\parskip}{6pt plus 2pt minus 1pt}}
}{% if KOMA class
  \KOMAoptions{parskip=half}}
\makeatother
\usepackage{xcolor}
\usepackage[margin=0.85in]{geometry}
\usepackage{longtable,booktabs,array}
\usepackage{calc} % for calculating minipage widths
\usepackage{etoolbox}
\makeatletter
\patchcmd\longtable{\par}{\if@noskipsec\mbox{}\fi\par}{}{}
\makeatother
\IfFileExists{footnotehyper.sty}{\usepackage{footnotehyper}}{\usepackage{footnote}}
\makesavenoteenv{longtable}
\usepackage{graphicx}
\makeatletter
\def\maxwidth{\ifdim\Gin@nat@width>\linewidth\linewidth\else\Gin@nat@width\fi}
\def\maxheight{\ifdim\Gin@nat@height>\textheight\textheight\else\Gin@nat@height\fi}
\makeatother
\setkeys{Gin}{width=\maxwidth,height=\maxheight,keepaspectratio}
\makeatletter
\def\fps@figure{htbp}
\makeatother
\ifLuaTeX
  \usepackage{selnolig}  % disable illegal ligatures
\fi
\usepackage{bookmark}
\IfFileExists{xurl.sty}{\usepackage{xurl}}{} % add URL line breaks if available
\hypersetup{
  pdftitle={Synaptic Dendritic Neuron Model (scDNM) for Cell Type Classification in Single-Cell RNA Sequencing Data},
  hidelinks,
  pdfcreator={LaTeX via pandoc}}

\title{Synaptic Dendritic Neuron Model (scDNM) for Cell Type Classification in Single-Cell RNA Sequencing Data}
\author{}
\date{}

\begin{document}
\maketitle

%\emph{Comparative evaluation across six binary classification tasks using leakage-free nested cross-validation}

\begin{center}

{\large\bfseries
Deniz Karap{\i}nar\textsuperscript{1},
Gaurav Choudhary\textsuperscript{2,3},
Melih Agraz\textsuperscript{2,3,4,*}
\par}

\vspace{0.5cm}

\small

\textsuperscript{1}
Department of Management and Organization,
Alucra Turan Bulut{\c c}u Vocational School,
Giresun University, Giresun, T\"urkiye.

\vspace{0.2cm}

\textsuperscript{2}
Division of Cardiology,
Brown University Health,
Providence, RI, USA.

\vspace{0.2cm}

\textsuperscript{3}
Department of Medicine,
The Warren Alpert Medical School of Brown University,
Providence, RI, USA.

\vspace{0.2cm}

\textsuperscript{4}
Department of Data Science and Analytics,
Giresun University, Giresun, T\"urkiye.

\vspace{0.3cm}

\textsuperscript{*}\textbf{Corresponding author:}
Melih Agraz\\
\texttt{melih\_agraz@brown.edu}

\end{center}

\section{Abstract}\label{abstract}

Objective. Cell type classification in single-cell RNA sequencing (scRNA-seq) data is challenging because of high dimensionality, sparsity, and transcriptionally similar cell types. This study develops scDNM, an adaptation of the dendritic neuron model (DNM) to scRNA-seq data, and compares it with conventional classifiers within a leakage-free evaluation framework.

Methods. scDNM comprises a synapse layer that models each gene--dendrite connection as a learnable soft threshold, a supervised gene pool selected from training data, staged training (first the dendritic model and then a residual global linear pathway), and an ensemble. Six binary tasks involving biologically similar and dissimilar cell types were constructed from the Paul15, PBMC3k, and Nestorowa datasets. For all models, highly variable gene selection, scaling, and hyperparameter selection were performed independently of the outer test fold using inner three-fold cross-validation with log-loss as the selection criterion.

Results. scDNM ranked first in three of six binary tasks (tied with SVM in one), second on Nestorowa, and achieved the best mean accuracy rank (2.00; XGBoost 2.75, Random Forest 3.33, logistic regression 3.33, and SVM 3.58). It also ranked first on average for AUC and F1. In an additional six-class PBMC3k experiment, scDNM achieved 95.997\% accuracy and 0.944 macro-F1, compared with 94.243\% and 0.914, respectively, for the official SingleR implementation. Official Garnett classifiers returned only ``Unknown'' labels in the tested environment and were therefore treated as a zero-coverage technical outcome rather than ranked as ordinary predictions. Ablation analysis showed reduced accuracy when dendrites or the synapse layer were removed. However, scDNM trailed tree-based models by approximately seven percentage points on the 14Mo--13Baso task and had lower mean binary-task accuracy (95.32\%) than XGBoost (96.20\%).

Conclusion. The dendritic architecture based on synaptic thresholds is competitive with strong conventional models for scRNA-seq cell type classification, ranking first or second in most tasks. Across-task rank differences were not statistically significant with six tasks; broader datasets are needed to test claims of general superiority.

\textbf{Keywords: scRNA-seq, cell type classification, dendritic neuron model, synapse layer, nested cross-validation, machine learning}

\section{1 Introduction}\label{introduction}

Single-cell RNA sequencing technologies have enabled the investigation of cellular diversity within tissues at the gene-expression level. However, single-cell data are high-dimensional, sparse, and noisy. Biological heterogeneity, transcriptionally similar cell types, unequal representation of cell types across datasets, and differences between training and target data further complicate classification. Consequently, cell type classification requires not only an appropriate classifier but also the correct placement of feature selection and scaling within the evaluation design.

Clustering and classification address related but distinct biological questions. Unsupervised clustering asks which cells resemble one another in the observed dataset and can reveal previously unrecognized states or subpopulations. A cluster is not automatically equivalent to a cell type: several clusters may represent different states of the same lineage, while a single cluster may contain unresolved identities. Supervised classification instead asks which predefined biological category best describes each cell, using labeled reference cells or externally defined marker information. It can therefore assign individual cells without requiring them to be grouped into clusters first.

Reliable classification is important when established cell identities must be transferred consistently to new samples, donors, or experiments. It supports reproducible annotation of large datasets, quantification of known populations, separation of transcriptionally similar subtypes, detection of rare populations, and comparison of cell-type composition across biological conditions. The central methodological challenge is to learn decision rules that remain accurate despite thousands of sparse gene measurements, limited labeled cells, class imbalance, and overlapping expression programs.

To address these challenges, numerous approaches have been developed based on support vector machines (SVMs), probabilistic models, ensemble learning, hierarchical classification, and deep learning. Comparative studies nevertheless indicate that performance depends on both the algorithm and dataset characteristics. A comparison of 22 classification methods across 27 scRNA-seq datasets found that a general-purpose linear-kernel SVM performed best overall \cite{1}. The effects of classifiers, feature selection, and reference construction have also been investigated \cite{2}, while another comparison of supervised machine learning algorithms demonstrated that performance depends on the dataset and evaluation conditions \cite{3}. In a more recent comparison, an RBF-kernel SVM performed best on three of four datasets \cite{4}, and the choice of SVM kernel was also shown to influence classification performance \cite{5}.

Some methodological developments rely on probabilistic and statistical modeling. scPred \cite{6} combines principal-component-based dimensionality reduction with probabilistic predictions from an SVM, whereas SciBet \cite{7} integrates discriminative gene selection with probabilistic classification based on a multinomial model. A probabilistic gene-expression barcode has been used for cell type prediction while addressing technical variation between studies \cite{8}. PCLDA \cite{9} combines t-test-based gene selection with principal component analysis (PCA) and linear discriminant analysis (LDA). Methods exploiting hierarchical relationships among cell types are also widespread. CHETAH \cite{10} uses hierarchical classification and intermediate or unassigned categories for uncertain predictions; scClassify \cite{11} combines hierarchical classification with ensemble learning. scAnnotatR \cite{12} organizes SVM classifiers hierarchically to support prediction at different levels of cell type granularity and the handling of unknown cells, while ImmClassifier \cite{13} combines cell ontology and classifiers trained on different references to classify immune-cell subtypes. A hierarchical cross-entropy loss incorporating cell type hierarchy into model training was reported to improve macro-F1 by 12--15\% on out-of-distribution test data from unseen studies \cite{14}.

Two established annotation tools illustrate complementary strategies. SingleR assigns labels by comparing each query cell with labeled reference profiles using correlation-based scores over discriminative genes \cite{43}. Garnett uses user-defined marker rules to select representative cells and trains hierarchical elastic-net multinomial regression models to classify cells, with an explicit ``Unknown'' outcome for insufficiently confident assignments \cite{44}. In contrast, scDNM learns gene-level soft thresholds and nonlinear combinations of those thresholded inputs across dendritic branches. Comparing these approaches therefore tests reference similarity, marker-informed regression, and learnable threshold interactions within a common PBMC setting.

Ensemble- and feature-selection-based approaches include Random Forest-based SingleCellNet \cite{15}, which classifies data across platforms and species; scIAE \cite{16}, which combines autoencoder-learned cell representations with ensemble classification; and QDE-SVM \cite{17}, which integrates quantum-inspired differential evolution for gene selection with SVM. Graph-based methods incorporate relationships between cell types and genes (gCAnno \cite{18}) or prior information from ligand--receptor interactions and intracellular pathways (scPriorGraph \cite{19}). Feature-fusion-based scMFF \cite{20} combines highly variable gene expression with power spectral density, PCA, and graph neural network representations. Among deep learning approaches, ItClust \cite{21} combines transfer learning with clustering and cell type classification and also addresses new cell types in target data. scDeepInsight \cite{22} transforms gene-expression profiles into image-like representations for convolutional neural network classification and reports 87.5\% mean accuracy across six independent PBMC test datasets. scTrans \cite{23} learns cell representations using a sparse-attention Transformer and contrastive pretraining, whereas scTab \cite{24}, developed on a compilation of 22.2 million cells, achieved higher macro-F1 than XGBoost and MLP on donor-separated test data. In low-label settings, Random Forest was reported to distinguish similar T-cell subtypes more successfully than logistic regression when only 10\% of PBMC3k cells were labeled \cite{25}.

Overall, the literature shows that scRNA-seq cell type classification cannot be reduced to choosing a single algorithm; performance is associated with feature representation, reference-data quality, cell type granularity, cell counts per class, and differences between training and target data \cite{1,2,3}. scDNM-VAE has employed signed dendritic gating for deep clustering of scRNA-seq data \cite{42}. Dendritic neuron models are learning models that represent nonlinear interactions among synaptic inputs on dendritic branches and have been used for classification, approximation, and prediction \cite{26,27}. Because this structure allows related gene groups to be processed in separate dendrites, it merits investigation for high-dimensional scRNA-seq problems with relatively small sample sizes.

In the classical DNM, each synapse is modeled as an excitatory or inhibitory connection by passing its input through a sigmoid threshold function; dendrites combine synaptic outputs, the membrane layer aggregates dendrites, and the soma produces the final decision \cite{27}. In scRNA-seq data, whether a gene is ``on'' or ``off'' in a particular cell type is often discriminative; gene-level thresholds are therefore a natural modeling unit. One reason tree-based methods are effective on such data is their ability to learn gene-level thresholds directly.

Related biomedical classification studies further illustrate methodological challenges relevant to single-cell transcriptomic analysis. Semi-supervised learning has improved thyroid disease prediction by exploiting unlabeled observations when fully annotated clinical data are limited \cite{45}. In high-dimensional breast cancer microarray data, feature-selection strategies have been shown to affect predictive performance and the identification of informative genes \cite{46}.  Related work on RNA-seq analysis has demonstrated the
utility of integrating autoencoders, data augmentation,
and feature selection in machine learning-based genomic
analysis \cite{48}. Such approaches highlight the
importance of informative feature representations when
analyzing high-dimensional transcriptomic data.
 More directly related to the present architecture, a hybrid model combining a dendritic neuron model with logistic regression achieved competitive performance in diabetes prediction, supporting the application of dendritic computation to biomedical classification \cite{47}. Collectively, these studies motivate computational frameworks that can accommodate limited labels, high-dimensional biological measurements, and nonlinear interactions while retaining clinically or biologically interpretable decision components.

In this study, we developed scDNM, adapting the synapse--dendrite--soma structure of the classical DNM to scRNA-seq data. The contributions are: (i) combining a synapse layer, in which each gene--dendrite connection is a learnable soft threshold, with a training-data-derived gene pool and staged residual training; (ii) an evaluation framework in which preprocessing, feature selection, hyperparameter tuning, and gene-pool selection are fully independent of the outer test fold; (iii) comparisons with five conventional classifiers and mean-rank analysis on six binary tasks involving biologically similar and dissimilar cell types from three datasets; (iv) an additional six-class PBMC3k comparison with the official SingleR and Garnett packages; and (v) an ablation analysis quantifying the contributions of model components.

\section{2 Materials and Methods}\label{materials-and-methods}

\subsection{2.1 Datasets and classification tasks}\label{datasets-and-classification-tasks}

Six binary tasks were constructed from three datasets (Table 1). Paul15 contains mouse myeloid progenitor cells \cite{28}; PBMC3k contains peripheral blood mononuclear cells from a healthy donor profiled using the 10x Genomics Chromium platform \cite{29,30}; and Nestorowa contains mouse hematopoietic stem and progenitor cells \cite{31}. These publicly available and widely used datasets were selected to span two species, peripheral and progenitor compartments, different feature dimensions, different class sizes, and a range of biological separation. They are established reference datasets in single-cell method development, but the particular binary cell-type pairs defined here should not be interpreted as a universally standardized benchmark suite. The pairs were specified to cover both transcriptionally similar types (e.g., 14Mo--13Baso and CD4 T--CD8 T) and clearly distinct types (e.g., 3Ery--16Neu and CD4 T--CD14+ monocytes), thereby testing performance across varying classification difficulty. Pair selection was based on biological relationships and data availability rather than observed model performance. Paul15 raw counts were obtained through Scanpy \cite{32}. For PBMC3k, raw counts were reconstructed from the log-normalized matrix in the Scanpy tutorial data object (by dividing the per-cell inverse exponential transformation by the smallest nonzero value); the maximum deviation of reconstructed values from integers was $6\cdot 10^{-5}$, and cell totals exactly matched the original total counts. In Nestorowa, cells labeled HSPC and Prog were included and LT-HSC cells were excluded. The Nestorowa dataset was generated from 12-week-old female C57BL/6 mice. Animal sex information was not available in the accessed Paul15 dataset, and sex-stratified analysis was not applicable to the present secondary computational comparison. Genes not expressed in any cell were removed; this step does not use labels.

\begin{longtable}[]{@{}
  >{\raggedright\arraybackslash}p{(\columnwidth - 10\tabcolsep) * \real{0.2976}}
  >{\raggedright\arraybackslash}p{(\columnwidth - 10\tabcolsep) * \real{0.1845}}
  >{\raggedright\arraybackslash}p{(\columnwidth - 10\tabcolsep) * \real{0.1845}}
  >{\raggedright\arraybackslash}p{(\columnwidth - 10\tabcolsep) * \real{0.1071}}
  >{\raggedright\arraybackslash}p{(\columnwidth - 10\tabcolsep) * \real{0.1190}}
  >{\raggedright\arraybackslash}p{(\columnwidth - 10\tabcolsep) * \real{0.1071}}@{}}
\toprule\noalign{}
\begin{minipage}[b]{\linewidth}\raggedright
Task
\end{minipage} & \begin{minipage}[b]{\linewidth}\raggedright
Positive class
\end{minipage} & \begin{minipage}[b]{\linewidth}\raggedright
Negative class
\end{minipage} & \begin{minipage}[b]{\linewidth}\raggedright
Cells
\end{minipage} & \begin{minipage}[b]{\linewidth}\raggedright
Genes
\end{minipage} & \begin{minipage}[b]{\linewidth}\raggedright
Difficulty
\end{minipage} \\
\midrule\noalign{}
\endhead
\bottomrule\noalign{}
\endlastfoot
Paul15 7MEP--10GMP & 7MEP (167) & 10GMP (153) & 320 & 3.451 & moderate \\
Paul15 14Mo--13Baso & 14Mo (373) & 13Baso (300) & 673 & 3.416 & similar \\
Paul15 3Ery--16Neu & 3Ery (246) & 16Neu (164) & 410 & 3.409 & distinct \\
PBMC3k CD4 T--CD8 T & CD8 T (316) & CD4 T (1.144) & 1.460 & 13.275 & similar \\
PBMC3k CD4 T--CD14+ Mono & CD14+ Mono (480) & CD4 T (1.144) & 1.624 & 13.511 & distinct \\
Nestorowa HSPC--Prog & Prog (852) & HSPC (852) & 1.704 & 43.103 & moderate \\
\end{longtable}

\textbf{Table 1. Classification tasks. The gene count denotes genes expressed in at least one cell.}

\subsection{2.2 Preprocessing and prevention of data leakage}\label{preprocessing-and-prevention-of-data-leakage}

Within each training partition (outer or inner training), F highly variable genes (HVGs) were first selected from raw counts using Seurat v3 \cite{33}. Total expression was then normalized to 10,000 per cell and log1p-transformed:

\(x_{ig}\  = \ \ln\left( 1\  + \ 10^{4} \cdot \frac{r_{ig}}{\sum_{g'}^{}r_{ig'}} \right)\) (1)

Here, $r_{ig}$ is the raw count for gene g in cell i. For each selected gene, the mean and standard deviation were computed using training cells only, and the same transformation was applied unchanged to test cells:

\({x\tilde{}}_{if}\  = \ \frac{x_{if}\  - \ \mu_{f}^{eğt}}{s_{f}^{eğt}}\) (2)

HVG selection, scaling, and the supervised scDNM gene pool were reconstructed in every training partition. Thus, cells in the outer test fold did not contribute to any feature selection, scaling, hyperparameter, or gene-pool decisions.

\subsection{2.3 scDNM architecture}\label{scdnm-architecture}

scDNM adapts the four layers of the classical DNM \cite{26,27} to scRNA-seq data (Figure 1). Synapse layer: for gene f connected to dendrite j, the synaptic output is a learnable soft threshold applied to standardized gene expression:

\(s_{fj}(x)\  = \ \sigma\left( k\left( w_{fj}{x\tilde{}}_{f}\  - \ q_{fj} \right) \right)\) (3)

\emph{The sign of w\_fj determines whether the synapse is excitatory or inhibitory; q\_fj controls the threshold location and k controls its sharpness. Thresholds are initialized at the midpoint between the ``not expressed'' level and the mean expression among expressing cells in the training set, so each synapse begins training as a gene ``on/off'' detector. Dendrite layer: each dendrite computes a weighted sum of synaptic outputs over its gene subset S\_j:}

\(z_{j}(x)\  = \ \sigma\left( \sum_{f\  \in \ S_{j}}^{}{a_{fj}s_{fj}(x)}\  + \ b_{j} \right)\) (4)

In the classical DNM, dendritic outputs are products of synaptic outputs. Because products across hundreds of genes numerically collapse toward zero, an additive dendrite was used. Membrane and soma: dendritic outputs are combined using learnable weights v\_j, a global linear pathway is added, and the result is converted into a probability using a sigmoid:

\(p\left( y\  = \ 1\ |\ x \right)\  = \ \sigma\left( \sum_{j = 1}^{K}{v_{j}z_{j}(x)}\  + \ {x\tilde{}}^{\top}w_{g}\  + \ b \right)\) (5)

\textbf{Supervised gene pool: dendritic subsets are randomly drawn from the top P genes ranked by the absolute standardized difference in class means computed from inner-training cells:}

\(\Delta_{f}\  = \ \frac{1}{n_{1}}\sum_{i:\ y_{i}\  = \ 1}^{}{x\tilde{}}_{if}\  - \ \frac{1}{n_{0}}\sum_{i:\ y_{i}\  = \ 0}^{}{x\tilde{}}_{if}\) (6)

Each dendrite randomly selects m genes from the pool (random subspace), so dendrites process partially overlapping gene groups. In the configuration without a gene pool, dendrites draw m = 0.1\ensuremath{\cdot}F genes from all HVGs. Default settings are K = 16 dendrites, k = 4, P = 1000, and m = 0.2\ensuremath{\cdot}P.

\includegraphics[width=6.45833in,height=3.48958in]{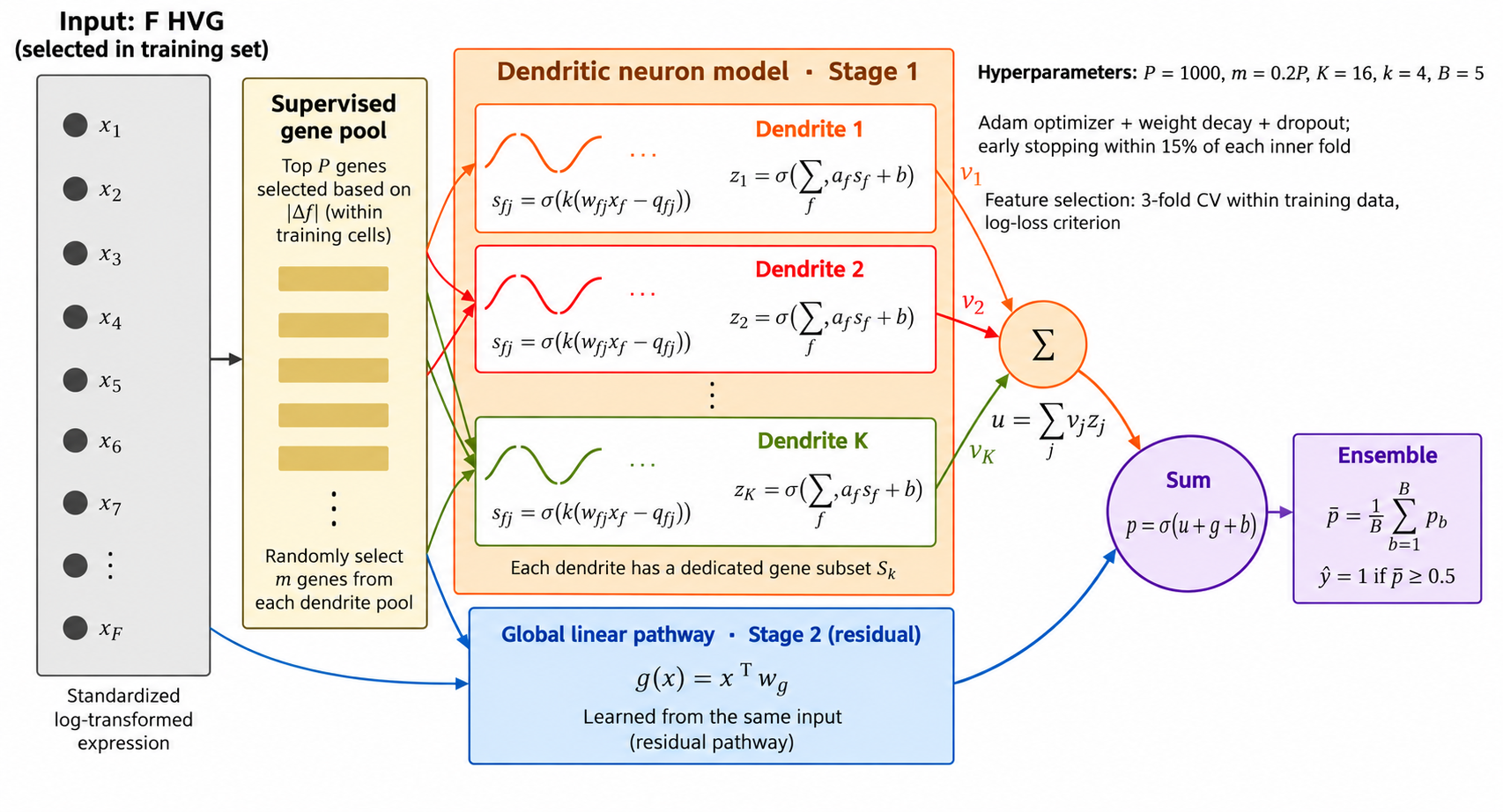}

\textbf{Figure 1. scDNM architecture. HVGs are selected within the training fold, and the supervised gene pool is constructed from training cells only. Each dendrite receives a random subset of genes from the pool; each gene--dendrite connection is a learnable soft threshold (synapse). In Stage 1, the dendritic model is trained; in Stage 2, the global linear pathway is trained on the residual while the dendritic logit is frozen. The final probability is the average of B networks.}

\subsection{2.4 Training}\label{training}

Training preserves the staged structure of scDNM. In Stage 1, only synapse, dendrite, membrane weights, and soma bias are trained. In Stage 2, dendritic parameters are frozen, and the global linear pathway learns residual information not explained by the dendritic logit. In both stages, binary cross-entropy loss was minimized with Adam \cite{34} (learning rate 0.01; batch size 128); weight decay of $10^{-3}$ and $10^{-2}$ was applied to dendritic parameters and the global pathway, respectively, and input dropout of 0.1 was used \cite{35}:

\(L\  = \  - \frac{1}{n}\sum_{i = 1}^{n}{y_{i}\ \ln\ p_{i}\  + \ \left( 1\  - \ y_{i} \right)\ \ln\left( 1\  - \ p_{i} \right)}\  + \ \lambda_{d}\|\theta_{d}\|^{2}\  + \ \lambda_{g}\| w_{g}\|^{2}\) (7)

In each stage, training was stopped if the loss on a stratified 15\% internal holdout from the training data did not improve for 25 epochs, and the best weights were restored (maximum 300 epochs). Probabilities from B = 5 networks trained with different seeds, gene subsets, and internal splits were averaged:

\(p\bar{}(x)\  = \ \frac{1}{B}\sum_{b = 1}^{B}p^{(b)}(x)\) (8)

\subsection{2.5 Comparison models and model selection}\label{comparison-models-and-model-selection}

scDNM was compared with logistic regression (LR), Random Forest (RF) \cite{36}, XGBoost \cite{37}, a radial basis function (RBF) kernel support vector machine (SVM), and a multilayer perceptron (MLP). Evaluation used nested cross-validation (Figure 2). Stratified five-fold cross-validation was used in the outer loop. Within each outer training set, stratified three-fold inner cross-validation selected the HVG count and model settings that minimized mean inner log-loss. The model was retrained on the entire outer training set using the selected configuration and evaluated once on the outer test fold. The decision threshold was fixed at 0.5 for all models. Candidate grids are presented in Table 2; the HVG grid was identical for all models within each task.

\begin{longtable}[]{@{}
  >{\raggedright\arraybackslash}p{(\columnwidth - 2\tabcolsep) * \real{0.2350}}
  >{\raggedright\arraybackslash}p{(\columnwidth - 2\tabcolsep) * \real{0.7650}}@{}}
\toprule\noalign{}
\begin{minipage}[b]{\linewidth}\raggedright
\textbf{Model}
\end{minipage} & \begin{minipage}[b]{\linewidth}\raggedright
Candidate settings
\end{minipage} \\
\midrule\noalign{}
\endhead
\bottomrule\noalign{}
\endlastfoot
HVG count & Paul15 7MEP--10GMP: \{500, 1000, 2000\}; other Paul15 and PBMC3k tasks: \{1000, 2000, 4000\} (limited by the available number of genes); Nestorowa: \{4000, 8000\} \\
scDNM & \{without pool, m = 0.1F\} and \{P = 1000, m = 0.2P\}; K = 16, k = 4, B = 5 \\
Logistic regression & C \ensuremath{\in} \{0.003; 0.01; 0.1\}* \\
Random Forest & 300 trees; (max\_features, min\_samples\_leaf) \ensuremath{\in} \{($\sqrt{p}$, 1), ($\sqrt{p}$, 3), (0.2, 1)\}* \\
XGBoost & 300 trees, learning rate 0.05, subsampling 0.8; depth \ensuremath{\in} \{2, 3, 4\}* \\
SVM (RBF) & C \ensuremath{\in} \{0.5; 1; 5\}*, \ensuremath{\gamma} = scale \\
MLP & \{(100), \ensuremath{\alpha} = $10^{-3}$\}, \{(100), \ensuremath{\alpha} = $10^{-1}$\}, \{(50, 25), \ensuremath{\alpha} = $10^{-2}$\}; early stopping \\
\end{longtable}

\textbf{Table 2. Candidate grids for model selection. *For tasks other than Paul15 7MEP--10GMP, only two candidates were evaluated due to computational constraints: LR C \ensuremath{\in} \{0.003, 0.01\}, RF \{($\sqrt{p}$, 1), (0.2, 1)\}, XGBoost depth \ensuremath{\in} \{3, 4\}, and SVM C \ensuremath{\in} \{1, 5\}.}

\includegraphics[width=6.45833in,height=2.63542in]{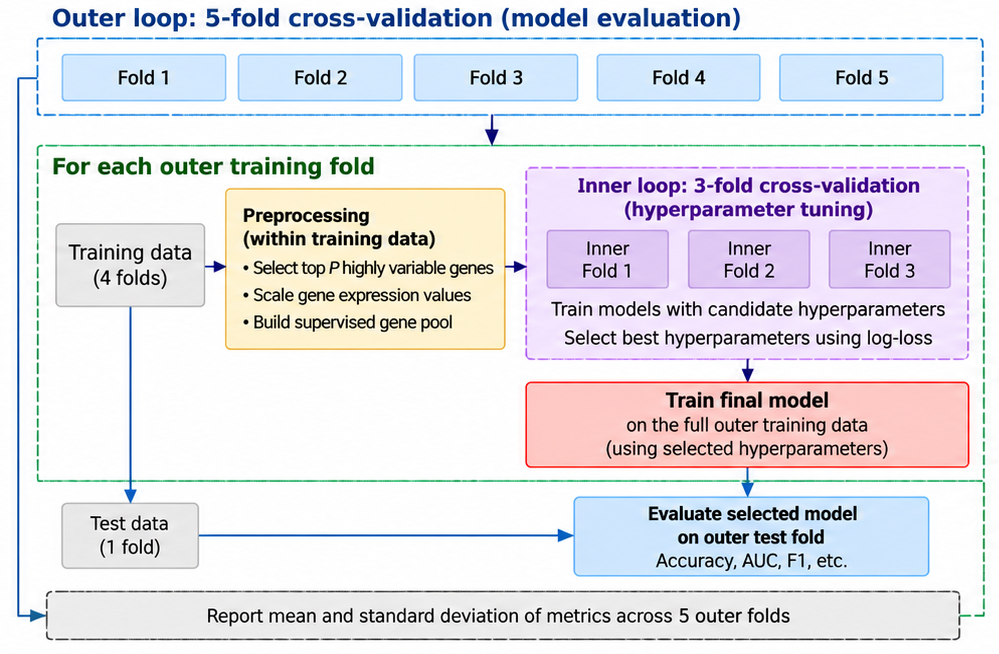}

\textbf{Figure 2. Leakage-free nested cross-validation. The outer test fold is excluded from all selection steps; HVG selection, scaling, and the gene pool are fitted using only the relevant training cells.}

Outer splits were repeated three times for Paul15 7MEP--10GMP (seeds 2026, 7, 123), twice for Nestorowa (seeds 11, 99), and once for the remaining four tasks (seed 2026). Architectural decisions for scDNM were made exclusively using a development subset drawn from the training data of a different Nestorowa split (seed 42); outer evaluation splits were not used during development, and neither Paul15 nor PBMC3k tasks were used for development.

\subsection{2.6 Evaluation metrics and statistical analysis}\label{evaluation-metrics-and-statistical-analysis}

Accuracy, F1, and area under the receiver operating characteristic curve (AUC) were computed for each outer test fold and reported as the mean \ensuremath{\pm} standard deviation across folds. Within-task comparisons used two-sided Wilcoxon signed-rank tests on fold-level results \cite{38}. For across-task comparisons, models were ranked by mean accuracy within each task (average ranks for ties); mean ranks were calculated and the Friedman test was applied \cite{39}:

\({R\bar{}}_{m}\  = \ \frac{1}{T}\sum_{t = 1}^{T}r_{tm}\) (9)

Here, r\_tm denotes the rank of model m on task t, and T = 6 is the number of tasks. Preprocessing was implemented with Scanpy \cite{32}, conventional models with scikit-learn \cite{40} and XGBoost \cite{37}, and scDNM with PyTorch \cite{41}.

\subsection{2.7 Multiclass PBMC3k annotation experiment}\label{multiclass-pbmc3k-annotation}

An additional experiment evaluated six-class annotation on PBMC3k. CD14+ and FCGR3A+ monocytes were merged as Monocytes, and the 15 megakaryocytes were excluded because the official Garnett PBMC marker definition did not contain that class. The resulting 2,623 cells comprised B cells, CD4 T cells, CD8 T cells, dendritic cells, monocytes, and NK cells. A stratified five-fold split with seed 2026 was shared by all methods. scDNM used one-versus-rest fitting without changing the binary learner. SingleR version 2.14.2 was run through the official Bioconductor package \cite{43}; within each outer fold, the training cells formed the reference and the held-out cells formed the test set, with marker selection confined to the training reference. Accuracy, balanced accuracy, macro-F1, and weighted-F1 were calculated on each held-out fold.

Garnett version 0.1.24 was evaluated through the official R package and official human PBMC marker file \cite{44}. First, a classifier was trained within each outer training fold and applied to the corresponding test fold. Second, the developers' pretrained hsPBMC classifier (dated 2019-10-17) was applied. Cluster extension was disabled to prevent information sharing among held-out cells. The documented confidence ratio of 1.50 and sensitivity values of 1.10 and 1.01 were examined. Because Garnett can abstain by returning ``Unknown,'' specific-label coverage was reported separately; an all-Unknown output was treated as a technical zero-coverage result and not as an ordinary zero-accuracy classifier.

\section{3 Results}\label{results}

\subsection{3.1 Main comparison}\label{main-comparison}

Results across the six tasks are shown in Table 3 and Figure 3. scDNM ranked first on Paul15 7MEP--10GMP (98.23\%), PBMC3k CD4 T--CD8 T (95.07\%; tied with SVM), and PBMC3k CD4 T--CD14+ Mono (99.82\%), and second after XGBoost on Nestorowa (91.37\%). On 3Ery--16Neu, four of the five models tied at 99.76\% accuracy, while Random Forest classified all folds correctly. On 14Mo--13Baso, scDNM (87.67\%) outperformed logistic regression and SVM but had substantially lower accuracy than XGBoost (94.50\%) and Random Forest (94.36\%).

\begin{longtable}[]{@{}
  >{\raggedright\arraybackslash}p{(\columnwidth - 12\tabcolsep) * \real{0.1923}}
  >{\raggedright\arraybackslash}p{(\columnwidth - 12\tabcolsep) * \real{0.1346}}
  >{\raggedright\arraybackslash}p{(\columnwidth - 12\tabcolsep) * \real{0.1346}}
  >{\raggedright\arraybackslash}p{(\columnwidth - 12\tabcolsep) * \real{0.1346}}
  >{\raggedright\arraybackslash}p{(\columnwidth - 12\tabcolsep) * \real{0.1346}}
  >{\raggedright\arraybackslash}p{(\columnwidth - 12\tabcolsep) * \real{0.1346}}
  >{\raggedright\arraybackslash}p{(\columnwidth - 12\tabcolsep) * \real{0.1346}}@{}}
\toprule\noalign{}
\begin{minipage}[b]{\linewidth}\raggedright
Task (fold)
\end{minipage} & \begin{minipage}[b]{\linewidth}\raggedright
\textbf{scDNM}
\end{minipage} & \begin{minipage}[b]{\linewidth}\raggedright
\textbf{XGBoost}
\end{minipage} & \begin{minipage}[b]{\linewidth}\raggedright
\textbf{RF}
\end{minipage} & \begin{minipage}[b]{\linewidth}\raggedright
\textbf{LR}
\end{minipage} & \begin{minipage}[b]{\linewidth}\raggedright
\textbf{SVM}
\end{minipage} & \begin{minipage}[b]{\linewidth}\raggedright
\textbf{MLP}
\end{minipage} \\
\midrule\noalign{}
\endhead
\bottomrule\noalign{}
\endlastfoot
Paul15 7MEP--10GMP (15) & \begin{minipage}[t]{\linewidth}\raggedright
98,23\\
\ensuremath{\pm} 1,43 \ensuremath{\cdot} rank 1

\ensuremath{\pm} 1,43 \ensuremath{\cdot} sıra 1\strut
\end{minipage} & \begin{minipage}[t]{\linewidth}\raggedright
96,15\\
\ensuremath{\pm} 2,28 \ensuremath{\cdot} rank 4

\ensuremath{\pm} 2,28 \ensuremath{\cdot} sıra 4\strut
\end{minipage} & \begin{minipage}[t]{\linewidth}\raggedright
95,21\\
\ensuremath{\pm} 3,31 \ensuremath{\cdot} rank 5

\ensuremath{\pm} 3,31 \ensuremath{\cdot} sıra 5\strut
\end{minipage} & \begin{minipage}[t]{\linewidth}\raggedright
97,19\\
\ensuremath{\pm} 1,89 \ensuremath{\cdot} rank 2

\ensuremath{\pm} 1,89 \ensuremath{\cdot} sıra 2\strut
\end{minipage} & \begin{minipage}[t]{\linewidth}\raggedright
96,25\\
\ensuremath{\pm} 1,54 \ensuremath{\cdot} rank 3

\ensuremath{\pm} 1,54 \ensuremath{\cdot} sıra 3\strut
\end{minipage} & 92,81

\ensuremath{\pm} 4,33 \\
Paul15 14Mo--13Baso (5) & \begin{minipage}[t]{\linewidth}\raggedright
87,67\\
\ensuremath{\pm} 3,72 \ensuremath{\cdot} rank 3

\ensuremath{\pm} 3,72 \ensuremath{\cdot} sıra 3\strut
\end{minipage} & \begin{minipage}[t]{\linewidth}\raggedright
94,50\\
\ensuremath{\pm} 1,12 \ensuremath{\cdot} rank 1

\ensuremath{\pm} 1,12 \ensuremath{\cdot} sıra 1\strut
\end{minipage} & \begin{minipage}[t]{\linewidth}\raggedright
94,36\\
\ensuremath{\pm} 2,43 \ensuremath{\cdot} rank 2

\ensuremath{\pm} 2,43 \ensuremath{\cdot} sıra 2\strut
\end{minipage} & \begin{minipage}[t]{\linewidth}\raggedright
82,76\\
\ensuremath{\pm} 2,07 \ensuremath{\cdot} rank 5

\ensuremath{\pm} 2,07 \ensuremath{\cdot} sıra 5\strut
\end{minipage} & \begin{minipage}[t]{\linewidth}\raggedright
84,10\\
\ensuremath{\pm} 2,00 \ensuremath{\cdot} rank 4

\ensuremath{\pm} 2,00 \ensuremath{\cdot} sıra 4\strut
\end{minipage} & 76,67

\ensuremath{\pm} 1,40 \\
Paul15 3Ery--16Neu (5) & \begin{minipage}[t]{\linewidth}\raggedright
99,76\\
\ensuremath{\pm} 0,55 \ensuremath{\cdot} rank 3,5

\ensuremath{\pm} 0,55 \ensuremath{\cdot} sıra 3,5\strut
\end{minipage} & \begin{minipage}[t]{\linewidth}\raggedright
99,76\\
\ensuremath{\pm} 0,55 \ensuremath{\cdot} rank 3,5

\ensuremath{\pm} 0,55 \ensuremath{\cdot} sıra 3,5\strut
\end{minipage} & \begin{minipage}[t]{\linewidth}\raggedright
100,00\\
\ensuremath{\pm} 0,00 \ensuremath{\cdot} rank 1

\ensuremath{\pm} 0,00 \ensuremath{\cdot} sıra 1\strut
\end{minipage} & \begin{minipage}[t]{\linewidth}\raggedright
99,76\\
\ensuremath{\pm} 0,55 \ensuremath{\cdot} rank 3,5

\ensuremath{\pm} 0,55 \ensuremath{\cdot} sıra 3,5\strut
\end{minipage} & \begin{minipage}[t]{\linewidth}\raggedright
99,76\\
\ensuremath{\pm} 0,55 \ensuremath{\cdot} rank 3,5

\ensuremath{\pm} 0,55 \ensuremath{\cdot} sıra 3,5\strut
\end{minipage} & 96,83

\ensuremath{\pm} 1,64 \\
PBMC3k CD4 T--CD8 T (5) & \begin{minipage}[t]{\linewidth}\raggedright
95,07\\
\ensuremath{\pm} 1,92 \ensuremath{\cdot} rank 1,5

\ensuremath{\pm} 1,92 \ensuremath{\cdot} sıra 1,5\strut
\end{minipage} & \begin{minipage}[t]{\linewidth}\raggedright
94,38\\
\ensuremath{\pm} 1,58 \ensuremath{\cdot} rank 4,5

\ensuremath{\pm} 1,58 \ensuremath{\cdot} sıra 4,5\strut
\end{minipage} & \begin{minipage}[t]{\linewidth}\raggedright
94,38\\
\ensuremath{\pm} 1,92 \ensuremath{\cdot} rank 4,5

\ensuremath{\pm} 1,92 \ensuremath{\cdot} sıra 4,5\strut
\end{minipage} & \begin{minipage}[t]{\linewidth}\raggedright
94,79\\
\ensuremath{\pm} 2,45 \ensuremath{\cdot} rank 3

\ensuremath{\pm} 2,45 \ensuremath{\cdot} sıra 3\strut
\end{minipage} & \begin{minipage}[t]{\linewidth}\raggedright
95,07\\
\ensuremath{\pm} 1,92 \ensuremath{\cdot} rank 1,5

\ensuremath{\pm} 1,92 \ensuremath{\cdot} sıra 1,5\strut
\end{minipage} & 92,19

\ensuremath{\pm} 2,90 \\
PBMC3k CD4 T--CD14+ Mono (5) & \begin{minipage}[t]{\linewidth}\raggedright
99,82\\
\ensuremath{\pm} 0,28 \ensuremath{\cdot} rank 1

\ensuremath{\pm} 0,28 \ensuremath{\cdot} sıra 1\strut
\end{minipage} & \begin{minipage}[t]{\linewidth}\raggedright
99,75\\
\ensuremath{\pm} 0,26 \ensuremath{\cdot} rank 2,5

\ensuremath{\pm} 0,26 \ensuremath{\cdot} sıra 2,5\strut
\end{minipage} & \begin{minipage}[t]{\linewidth}\raggedright
99,69\\
\ensuremath{\pm} 0,38 \ensuremath{\cdot} rank 4,5

\ensuremath{\pm} 0,38 \ensuremath{\cdot} sıra 4,5\strut
\end{minipage} & \begin{minipage}[t]{\linewidth}\raggedright
99,75\\
\ensuremath{\pm} 0,26 \ensuremath{\cdot} rank 2,5

\ensuremath{\pm} 0,26 \ensuremath{\cdot} sıra 2,5\strut
\end{minipage} & \begin{minipage}[t]{\linewidth}\raggedright
99,69\\
\ensuremath{\pm} 0,38 \ensuremath{\cdot} rank 4,5

\ensuremath{\pm} 0,38 \ensuremath{\cdot} sıra 4,5\strut
\end{minipage} & 99,32

\ensuremath{\pm} 0,51 \\
Nestorowa HSPC--Prog (10) & \begin{minipage}[t]{\linewidth}\raggedright
91,37\\
\ensuremath{\pm} 1,24 \ensuremath{\cdot} rank 2

\ensuremath{\pm} 1,24 \ensuremath{\cdot} sıra 2\strut
\end{minipage} & \begin{minipage}[t]{\linewidth}\raggedright
92,63\\
\ensuremath{\pm} 0,86 \ensuremath{\cdot} rank 1

\ensuremath{\pm} 0,86 \ensuremath{\cdot} sıra 1\strut
\end{minipage} & \begin{minipage}[t]{\linewidth}\raggedright
90,93\\
\ensuremath{\pm} 0,99 \ensuremath{\cdot} rank 3

\ensuremath{\pm} 0,99 \ensuremath{\cdot} sıra 3\strut
\end{minipage} & \begin{minipage}[t]{\linewidth}\raggedright
89,88\\
\ensuremath{\pm} 1,38 \ensuremath{\cdot} rank 4

\ensuremath{\pm} 1,38 \ensuremath{\cdot} sıra 4\strut
\end{minipage} & \begin{minipage}[t]{\linewidth}\raggedright
89,47\\
\ensuremath{\pm} 1,38 \ensuremath{\cdot} rank 5

\ensuremath{\pm} 1,38 \ensuremath{\cdot} sıra 5\strut
\end{minipage} & -- \\
Mean accuracy & 95,32 & 96,20 & 95,76 & 94,02 & 94,06 & -- \\
Mean rank & 2,00 & 2,75 & 3,33 & 3,33 & 3,58 & -- \\
\end{longtable}

\textbf{Table 3. Mean accuracy (\%) on outer test folds. The second line gives the standard deviation and within-task rank (1 = best; average ranks for ties; MLP was excluded from ranking because it was not evaluated on Nestorowa).}

\includegraphics[width=6.45833in,height=2.625in]{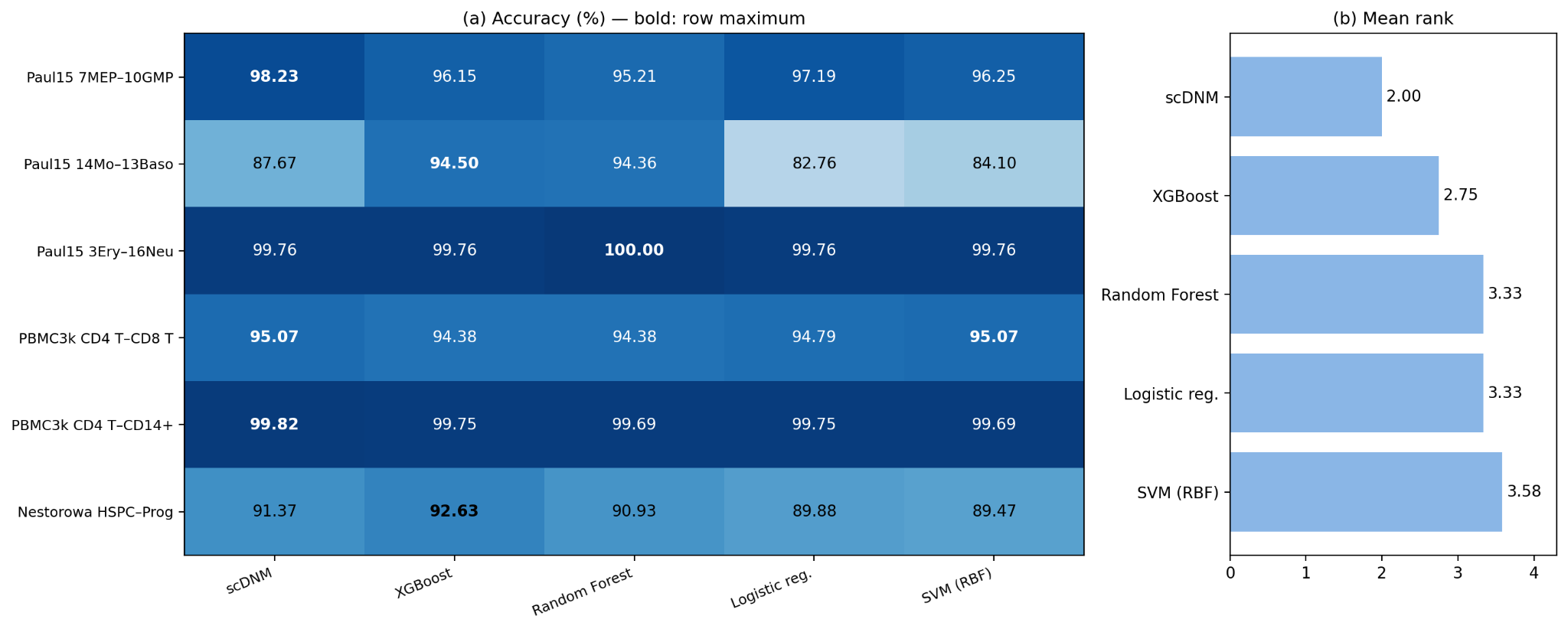}

\textbf{Figure 3. (a) Task-level accuracy and within-task rank; darker shading indicates better rank. (b) Mean rank across six tasks.}

scDNM had the best mean rank on all three metrics (Table 4): 2.00 for accuracy, 2.08 for AUC, and 2.17 for F1. XGBoost ranked second for accuracy (2.75). However, XGBoost (96.20\%) and Random Forest (95.76\%) exceeded scDNM (95.32\%) in mean accuracy, largely because of the 14Mo--13Baso task. The Friedman test was not significant across six tasks (p = 0.353 for accuracy).

\begin{longtable}[]{@{}
  >{\raggedright\arraybackslash}p{(\columnwidth - 12\tabcolsep) * \real{0.1667}}
  >{\raggedright\arraybackslash}p{(\columnwidth - 12\tabcolsep) * \real{0.1389}}
  >{\raggedright\arraybackslash}p{(\columnwidth - 12\tabcolsep) * \real{0.1389}}
  >{\raggedright\arraybackslash}p{(\columnwidth - 12\tabcolsep) * \real{0.1389}}
  >{\raggedright\arraybackslash}p{(\columnwidth - 12\tabcolsep) * \real{0.1389}}
  >{\raggedright\arraybackslash}p{(\columnwidth - 12\tabcolsep) * \real{0.1389}}
  >{\raggedright\arraybackslash}p{(\columnwidth - 12\tabcolsep) * \real{0.1389}}@{}}
\toprule\noalign{}
\begin{minipage}[b]{\linewidth}\raggedright
Metric
\end{minipage} & \begin{minipage}[b]{\linewidth}\raggedright
\textbf{scDNM}
\end{minipage} & \begin{minipage}[b]{\linewidth}\raggedright
\textbf{XGBoost}
\end{minipage} & \begin{minipage}[b]{\linewidth}\raggedright
\textbf{RF}
\end{minipage} & \begin{minipage}[b]{\linewidth}\raggedright
\textbf{LR}
\end{minipage} & \begin{minipage}[b]{\linewidth}\raggedright
\textbf{SVM}
\end{minipage} & \begin{minipage}[b]{\linewidth}\raggedright
\textbf{Friedman p}
\end{minipage} \\
\midrule\noalign{}
\endhead
\bottomrule\noalign{}
\endlastfoot
Accuracy & 2,00

\%95,32 & 2,75

\%96,20 & 3,33

\%95,76 & 3,33

\%94,02 & 3,58

\%94,06 & 0,353 \\
AUC & 2,08

\%98,13 & 3,17

\%98,82 & 3,33

\%98,28 & 2,83

\%97,32 & 3,58

\%97,07 & 0,417 \\
F1 & 2,17

\%94,32 & 2,50

\%94,99 & 3,25

\%94,56 & 3,33

\%93,01 & 3,75

\%93,07 & 0,378 \\
\end{longtable}

\textbf{Table 4. Mean ranks across six tasks (mean values shown on the second line). Lower ranks indicate better performance.}

\subsection{3.2 Paired comparisons}\label{paired-comparisons}

Fold-level paired differences are reported in Table 5. On Paul15 7MEP--10GMP, scDNM achieved higher accuracy than all competitors (uncorrected fold-level Wilcoxon p \ensuremath{\leq} 0.031). On Nestorowa, scDNM exceeded logistic regression and SVM on all 10 folds (p = 0.002), exceeded Random Forest on 8 of 10 folds (p = 0.119), and trailed XGBoost on every fold (\ensuremath{-}1.26 percentage points, p = 0.002).

\begin{longtable}[]{@{}
  >{\raggedright\arraybackslash}p{(\columnwidth - 8\tabcolsep) * \real{0.3162}}
  >{\raggedright\arraybackslash}p{(\columnwidth - 8\tabcolsep) * \real{0.1709}}
  >{\raggedright\arraybackslash}p{(\columnwidth - 8\tabcolsep) * \real{0.1709}}
  >{\raggedright\arraybackslash}p{(\columnwidth - 8\tabcolsep) * \real{0.1709}}
  >{\raggedright\arraybackslash}p{(\columnwidth - 8\tabcolsep) * \real{0.1709}}@{}}
\toprule\noalign{}
\begin{minipage}[b]{\linewidth}\raggedright
Task
\end{minipage} & \begin{minipage}[b]{\linewidth}\raggedright
\textbf{-- XGBoost}
\end{minipage} & \begin{minipage}[b]{\linewidth}\raggedright
\textbf{-- RF}
\end{minipage} & \begin{minipage}[b]{\linewidth}\raggedright
\textbf{-- LR}
\end{minipage} & \begin{minipage}[b]{\linewidth}\raggedright
\textbf{-- SVM}
\end{minipage} \\
\midrule\noalign{}
\endhead
\bottomrule\noalign{}
\endlastfoot
Paul15 7MEP--10GMP & +2,08

p = 0,009 & +3,02

p = 0,009 & +1,04

p = 0,031 & +1,98

p = 0,003 \\
Paul15 14Mo--13Baso & \ensuremath{-}6,83

p = 0,063 & \ensuremath{-}6,69

p = 0,125 & +4,91

p = 0,063 & +3,57

p = 0,125 \\
Paul15 3Ery--16Neu & +0,00

p = 1,000 & \ensuremath{-}0,24

p = 1,000 & +0,00

p = 1,000 & +0,00

p = 1,000 \\
PBMC3k CD4 T--CD8 T & +0,68

p = 0,563 & +0,68

p = 0,375 & +0,27

p = 0,750 & +0,00

p = 0,875 \\
PBMC3k CD4 T--CD14+ Mono & +0,06

p = 0,750 & +0,12

p = 0,500 & +0,06

p = 1,000 & +0,12

p = 0,500 \\
Nestorowa HSPC--Prog & \ensuremath{-}1,26

p = 0,002 & +0,44

p = 0,119 & +1,50

p = 0,002 & +1,91

p = 0,002 \\
\end{longtable}

\textbf{Table 5. Accuracy differences between scDNM and other models (percentage points; scDNM minus competitor) and fold-level Wilcoxon p-values.}

\subsection{3.3 Ablation}\label{ablation}

Component contributions were evaluated on five outer folds using a fixed configuration without inner selection (P = 1000, m = 0.2P; the HVG count was fixed to the most frequently selected value in the main comparison for each task; Table 6). The largest loss occurred when dendrites were removed: with only the global linear pathway, accuracy fell by 1.76 percentage points on average and decreased on 18 of 30 folds (p \textless{} 0.001). Removing the synapse layer, i.e., supplying unthresholded raw expression to dendrites, reduced accuracy by 1.22 points on average and decreased it on 19 of 30 folds (p \textless{} 0.001), particularly on 14Mo--13Baso (\ensuremath{-}4.31) and Paul15 7MEP--10GMP (\ensuremath{-}1.25). Removing the global pathway (\ensuremath{-}0.35, p = 0.155), removing the supervised gene pool (\ensuremath{-}0.28, p = 0.294), and using a single network (\ensuremath{-}0.53, p = 0.102) produced small, nonsignificant average losses; however, the global pathway made a substantial contribution on Paul15 7MEP--10GMP (\ensuremath{-}1.88), and the ensemble did so on 14Mo--13Baso (\ensuremath{-}2.37). These findings suggest that much of scDNM performance derives from its dendritic structure with synaptic thresholds.

\begin{longtable}[]{@{}
  >{\raggedright\arraybackslash}p{(\columnwidth - 14\tabcolsep) * \real{0.1987}}
  >{\raggedright\arraybackslash}p{(\columnwidth - 14\tabcolsep) * \real{0.1175}}
  >{\raggedright\arraybackslash}p{(\columnwidth - 14\tabcolsep) * \real{0.1175}}
  >{\raggedright\arraybackslash}p{(\columnwidth - 14\tabcolsep) * \real{0.1175}}
  >{\raggedright\arraybackslash}p{(\columnwidth - 14\tabcolsep) * \real{0.1175}}
  >{\raggedright\arraybackslash}p{(\columnwidth - 14\tabcolsep) * \real{0.1175}}
  >{\raggedright\arraybackslash}p{(\columnwidth - 14\tabcolsep) * \real{0.1175}}
  >{\raggedright\arraybackslash}p{(\columnwidth - 14\tabcolsep) * \real{0.0962}}@{}}
\toprule\noalign{}
\begin{minipage}[b]{\linewidth}\raggedright
Variant
\end{minipage} & \begin{minipage}[b]{\linewidth}\raggedright
\textbf{Paul15}

\textbf{7MEP--10GMP}
\end{minipage} & \begin{minipage}[b]{\linewidth}\raggedright
\textbf{Paul15}

\textbf{14Mo--13Baso}
\end{minipage} & \begin{minipage}[b]{\linewidth}\raggedright
\textbf{Paul15}

\textbf{3Ery--16Neu}
\end{minipage} & \begin{minipage}[b]{\linewidth}\raggedright
\textbf{PBMC3k}

\textbf{CD4 T--CD8 T}
\end{minipage} & \begin{minipage}[b]{\linewidth}\raggedright
\textbf{PBMC3k}

\textbf{CD4 T--CD14+ Mono}
\end{minipage} & \begin{minipage}[b]{\linewidth}\raggedright
\textbf{Nestorowa}

\textbf{HSPC--Prog}
\end{minipage} & \begin{minipage}[b]{\linewidth}\raggedright
Mean \ensuremath{\Delta}
\end{minipage} \\
\midrule\noalign{}
\endhead
\bottomrule\noalign{}
\endlastfoot
Full model & 98,44 & 88,86 & 100,00 & 95,00 & 99,75 & 91,20 & -- \\
No global pathway & 96,56

\ensuremath{-}1,88 & 88,26

\ensuremath{-}0,59 & 100,00

+0,00 & 95,27

+0,27 & 99,75

+0,00 & 91,32

+0,12 & \ensuremath{-}0,35 \\
No dendrites (global only) & 97,50

\ensuremath{-}0,94 & 83,06

\ensuremath{-}5,80 & 98,78

\ensuremath{-}1,22 & 93,56

\ensuremath{-}1,44 & 99,82

+0,06 & 89,96

\ensuremath{-}1,23 & \ensuremath{-}1,76 \\
No synapse layer & 97,19

\ensuremath{-}1,25 & 84,55

\ensuremath{-}4,31 & 99,76

\ensuremath{-}0,24 & 94,18

\ensuremath{-}0,82 & 99,69

\ensuremath{-}0,06 & 90,55

\ensuremath{-}0,65 & \ensuremath{-}1,22 \\
No supervised gene pool & 98,13

\ensuremath{-}0,31 & 87,51

\ensuremath{-}1,35 & 100,00

+0,00 & 95,00

+0,00 & 99,82

+0,06 & 91,14

\ensuremath{-}0,06 & \ensuremath{-}0,28 \\
Single network (no ensemble) & 97,81

\ensuremath{-}0,63 & 86,48

\ensuremath{-}2,37 & 100,00

+0,00 & 94,73

\ensuremath{-}0,27 & 99,88

+0,12 & 91,14

\ensuremath{-}0,06 & \ensuremath{-}0,53 \\
\end{longtable}

\textbf{Table 6. scDNM ablation: accuracy (\%), fixed configuration, five outer folds (second line: difference from full model, percentage points). The HVG count was fixed to the most frequent scDNM selection in the main comparison for each task; mean \ensuremath{\Delta} was computed over 30 folds. In the ``No dendrites'' variant, the soma bias was initialized from the training-set log-odds.}

\subsection{3.4 Selected configurations and computational cost}\label{selected-configurations-and-computational-cost}

Inner cross-validation selected the configuration with a supervised gene pool for scDNM in 32 of 45 outer folds. HVG counts generally favored the upper grid values (2000 in 14 of 15 folds for Paul15 7MEP--10GMP; 8000 in 9 of 10 folds for Nestorowa), whereas 1000 HVGs sufficed for PBMC3k CD4 T--CD14+ Mono. Similar trends were observed for all models, with more genes generally improving performance. Mean time per outer fold, including inner selection, was approximately 33 minutes for scDNM on Nestorowa and 4.5 minutes on Paul15 7MEP--10GMP (single CPU core), several times longer than XGBoost.

\subsection{3.5 Multiclass PBMC3k annotation}\label{multiclass-pbmc3k-results}

In the six-class PBMC3k experiment, scDNM outperformed the official SingleR implementation on all four aggregate metrics (Table 7). Mean accuracy was 0.960 for scDNM and 0.942 for SingleR, while macro-F1 was 0.944 and 0.914, respectively. The larger difference in balanced accuracy (0.940 versus 0.891) indicates that scDNM's advantage was not restricted to the most abundant classes.

\begin{table}[htbp]
\centering
\small
\begin{tabular}{lccccc}
\toprule
Method & Accuracy & Balanced accuracy & Macro-F1 & Weighted-F1 & Coverage \\
\midrule
scDNM-v4 OvR & $0.960 \pm 0.006$ & $0.940 \pm 0.022$ & $0.944 \pm 0.020$ & $0.960 \pm 0.006$ & 1.000 \\
SingleR 2.14.2 & $0.942 \pm 0.008$ & $0.891 \pm 0.021$ & $0.914 \pm 0.024$ & $0.941 \pm 0.008$ & 1.000 \\
\bottomrule
\end{tabular}
\caption{Six-class PBMC3k performance across five stratified outer folds (mean $\pm$ standard deviation). Coverage is the fraction receiving a specific class label.}
\label{tab:multiclass-pbmc}
\end{table}

Both the fold-trained and pretrained official Garnett classifiers returned ``Unknown'' for all 2,623 cells at confidence ratios 1.50, 1.10, and 1.01 (specific-label coverage = 0). The pretrained run also reported eight classifier genes absent from the input matrix and required on-the-fly conversion of an object created with an older igraph version. Because no specific prediction was produced, Garnett was not assigned accuracy or F1 values and was excluded from the numerical ranking; this outcome is reported as a reproducibility and software-compatibility limitation rather than evidence of inferior biological classification.

\section{4 Discussion}\label{discussion}

This study adapted the dendritic neuron model to scRNA-seq cell type classification and compared it with strong conventional models across six binary tasks. scDNM achieved the best mean rank for all three metrics, ranking first on three tasks and second on Nestorowa. These findings indicate competitive performance across a range of tasks.

The additional multiclass experiment provided a direct comparison with a widely used reference-based annotation method. scDNM exceeded official SingleR by 1.75 percentage points in accuracy and 3.00 points in macro-F1. This experiment used an internal reference formed from each outer training fold, so it evaluates within-dataset supervised annotation rather than transfer to an independent study. Accordingly, it supports multiclass feasibility but does not establish cross-study generalization.

In practical use, scDNM could be trained on a carefully annotated reference and then used to assign known identities to cells from a comparable target cohort. Its threshold-based structure may be particularly useful when closely related subtypes differ through combinations of marker genes turning on or off rather than through a single linear expression gradient. Beyond prediction, the learned synaptic thresholds and dendritic gene subsets provide candidate decision rules that can be examined against known marker biology. The method is therefore intended to complement, rather than replace, exploratory clustering: clustering can discover candidate states, after which classification can provide reproducible labels for predefined populations and support their quantification across samples.

Beyond transcriptomic data, deep learning approaches
have also been applied to ECG-based obstructive sleep
apnea classification \cite{49}. These applications
illustrate the broader relevance of learning informative
representations from complex biomedical measurements,
although their architectures and data characteristics
differ from those considered in the present study.
 The most important next validation step is a genuinely independent reference-to-query design. Future experiments should train on one donor cohort or study and test on another without target-specific feature selection, followed by cross-platform and cross-species evaluations where appropriate. Unknown-cell rejection should also be added and calibrated so that cells absent from the training taxonomy are not forced into known classes. Finally, atlas-scale testing, runtime reduction, and prospective biological review of the learned gene thresholds will determine whether the observed within-dataset advantage translates into a useful annotation workflow.

Importantly, mean rank evaluates the relative performance of classifiers across tasks rather than emphasizing absolute accuracy on individual datasets. The consistently favorable ranking of scDNM across accuracy, AUC, and F1 suggests that its performance is not driven solely by a single classification task, although the observed ranking differences were not statistically significant.

For a model containing dendritic components to be interpreted as ``dendritic,'' it is necessary to demonstrate that dendrites genuinely contribute to its decisions. The ablation analysis directly demonstrated this contribution: removing dendrites or the synapse layer reduced accuracy on a majority of the 30 folds. The synapse layer defines each gene--dendrite connection as a soft threshold and enables dendrites to combine gene-level on/off patterns. This architecture brings the threshold-based strength of tree models into a continuous, differentiable framework.

The 14Mo--13Baso task illustrates a limitation of the method. Here, tree-based models achieved approximately 7--12 percentage points higher accuracy than linear models and scDNM. Because monocytes and basophils may be distinguished by a small number of strong marker genes, trees may benefit from directly selecting sparse, sharp thresholds. Smaller gene pools, sharper synaptic thresholds, or sparsity-inducing regularization for gene selection could improve scDNM on such tasks; incorporating these options into the inner cross-validation grid warrants future investigation.

The evaluation design is another important aspect of these findings. Repeatedly using the same outer folds during model development or selecting from many candidates on a small validation subset may lead to optimistic performance estimates. Here, model-development and evaluation splits were separated, selection was performed by inner cross-validation using a smooth criterion such as log-loss, and the decision threshold was fixed. Evaluating all models on the same HVG grid and comparably sized candidate sets improved the fairness of the comparison.

scDNM dendrites can be viewed as combinations of thresholds defined over gene subsets. Synaptic parameters (w, q) indicate the expression level at which each gene excites or inhibits a dendrite. Systematically examining associations between dendritic outputs and known marker genes, and defining gene subsets using pathway databases such as KEGG or Reactome \cite{19}, could strengthen biological interpretability.

The main study is based on six binary tasks from three datasets, supplemented by one multiclass PBMC3k experiment; broader multiclass datasets such as Baron, Muraro, and Zheng68K were not investigated. Results from repeated cross-validation folds are not independent observations, and no multiple-testing correction was applied to fold-level Wilcoxon comparisons. Nominal p-values should therefore be regarded as exploratory rather than conclusive evidence of statistical superiority. With few tasks, across-task rank differences were not significant in the Friedman test. Tasks other than Paul15 7MEP--10GMP and Nestorowa were evaluated with only one repetition (five outer folds), limiting the power of fold-level tests (the smallest exact p-value with five folds is 0.0625). Because architectural decisions for scDNM were made using a development set drawn from Nestorowa training data, Nestorowa results should not be interpreted as completely independent validation. Due to computational constraints, candidate grids for some conventional models were reduced to two options on certain tasks. The Garnett experiment produced no specific labels under either fold-trained or official pretrained configurations; its outcome may reflect compatibility among the 2019 classifier object, current package dependencies, and incomplete model-gene overlap, and therefore cannot support a quantitative performance claim. Cross-platform and cross-species transfer \cite{15,21}, out-of-distribution generalization across studies \cite{14}, and atlas-scale cross-tissue classification \cite{24} remain outside the scope of this study.

\section{5 Conclusion}\label{conclusion}

scDNM, incorporating synaptic thresholds, a supervised gene pool, and staged residual training, achieved the best mean rank for accuracy, AUC, and F1 across six binary scRNA-seq classification tasks, ranking first on three tasks and second on Nestorowa. It also outperformed official SingleR in a six-class PBMC3k experiment. Ablation supported the contributions of dendrites and the synapse layer; however, the method trailed tree-based models on a task distinguished by a small number of strong markers. The findings support dendritic neuron models as a competitive option with potential for interpretability in scRNA-seq classification, while claims of general superiority require testing on broader, independent, and cross-study datasets.

\section{Author contributions}\label{author-contributions}

D.K. and M.A. contributed to conceptualization and study design. M.A. developed the methodology and software and performed the computational analyses and visualization. D.K. contributed to investigation, interpretation, and preparation of the original manuscript. G.C. contributed biological and clinical interpretation, supervision, and critical revision of the manuscript. All authors reviewed and approved the final manuscript. Author roles should be confirmed by all authors before submission.

\section{Data and code availability}\label{data-and-code-availability}

Paul15 data were obtained from Scanpy (scanpy.datasets.paul15), PBMC3k from the Scanpy tutorial data object, and Nestorowa from the Open Problems resource. Analysis code, official-package scripts, cell-level predictions, and fold-level results are available in the scdnm\_v2\_analiz folder.

\end{document}